# Reinforcement Learning-Based Joint Self-Optimisation Method for the Fuzzy Logic Handover Algorithm in 5G HetNets

Qianyu Liu[a], Chiew Foong Kwong[*, b], Sun Wei[c], Sijia, Zhou[c], Lincan Li[b] and Pushpendu Kar[d]

[a] International Doctoral Innovation Centre, University of Nottingham Ningbo China, Ningbo, China

[b] Department of Electrical and Electronic Engineering, University of Nottingham Ningbo China, Ningbo, China

[c] Department of Electrical and Electronic Engineering, University of Nottingham, Nottingham, UK

[d] School of Computer Science, University of Nottingham Ningbo China, Ningbo, China

* Corresponding Author: Chiew Foong Kwong (C.F.Kwong@nottingham.edu.cn)

**ABSTRACT**

5G heterogeneous networks (HetNets) can provide higher network coverage and system capacity to the user by deploying massive small base stations (BSs) within the 4G macro system. However, the large-scale deployment of small BSs significantly increases the complexity and workload of network maintenance and optimisation. The current handover (HO) triggering mechanism A3 event was designed only for mobility management in the macro system. Directly implementing A3 in 5G-HetNets may degrade the user mobility robustness. Motivated by the concept of self-organisation networks (SON), this study developed a self-optimised triggering mechanism to enable automated network maintenance and enhance user mobility robustness in 5G-HetNets. The proposed method integrates the advantages of subtractive clustering and Q-learning frameworks into the conventional fuzzy logic-based HO algorithm (FLHA). Subtractive clustering is first adopted to generate a membership function (MF) for the FLHA to enable FLHA with the self-configuration feature. Subsequently, Q-learning is utilised to learn the optimal HO policy from the environment as fuzzy rules that empower the FLHA with a self-optimisation function. The FLHA with SON functionality also overcomes the limitations of the conventional FLHA that must rely heavily on professional experience to design. The simulation results show that the proposed self-optimised FLHA can effectively generate MF and fuzzy rules for the FLHA. The proposed approach can minimise the HO, ping-pong HO, and HO failure ratios while improving network throughput and latency by comparing with conventional triggering mechanisms.

## 1. INTRODUCTION

With increasing user demand on high data rates and the Internet of things (IoT), the fifth-generation communications system (5G) has recently been commercialised. Heterogeneous networks (HetNets) have played a vital role in the deployment of 5G by routing massive small cells to the 4G macro system [1], [2]. This heterogeneous architecture allows a larger amount of simultaneous mobile data to be delivered by the small cells and offloads part of the data traffic load from the 4G macrocell. Therefore, the system capacity and network coverage are significantly improved compared to the 4G macrocell system. The massive deployment of small cells can also increase the complexity and workload of network maintenance. To reduce the capital and operational expenses of network maintenance, self-organisation networks (SON) have been defined by the Third-Generation Partnership Project (3GPP) to enable automated network operation while improving network performance [3]–[5]. Specifically, SON includes self-configuration, self-optimisation, and self-healing functionalities that aim to automate parameter configuration, parameter optimisation, and troubleshooting, respectively.

In the self-optimisation field, one of the most important use cases defined for radio access networks is mobility robustness optimisation [5]. During the movement of the user equipment (UE), the UE will frequently switch its connection with the base station (BS) to ensure seamless communication, which is known as the handover (HO) process. The HO process is triggered when the UE meets the entering condition of the A3 event [6], when the reference signal receiving power (RSRP) from the neighbouring BS is higher than serving the BS of the UE. The HO process can directly affect the user experience as it occurs during the data packet transmission. However, the current HO triggering mechanism A3 event is only considering one metric RSRP as decision criteria. This single metric mechanism is easily affected by interference and noise, resulting in several abnormal HO effects, such as unnecessary HOs, ping-pong HO, and HO failure [7]. Besides, these effects will become more severe in 5G-HetNets as the dense deployment of small BSs can introduce much stronger inter-cell interference. Thus, the main objective of mobility robustness optimisation is to reduce abnormal HO effects due to the triggering mechanism and increase network resources usage by minimising unnecessary HOs [8].

Based on our previous work in [9], this paper takes a step forward by synergising both Q-learning and subtractive clustering strengths to enable the fuzzy logic HO algorithm (FLHA) with self-optimisation functionality. The proposed self-optimised FLHA should enhance user mobility robustness by reducing the number of HOs, ping-pong effects, and HO failures while improving other networks KPIs, that is, network latency and throughput. Specifically, the contribution of this paper can be summarised as follows:

- First, we propose a self-optimised FLHA that can consider the multivariate analysis in uncertain environments. The fuzzy inference engine processes multiple network data to estimate the HO probabilities as HO triggering indicators.
- Second, we adopt the Q-learning framework to learn the optimal HO policy from the environment as fuzzy rules for the fuzzy inference system. This approach allows the FLHA to self-optimise its fuzzy rules by interacting with the environment.
- Finally, subtractive clustering is adopted to generate fuzzy sets for the fuzzy inference system to convert decision metrics into the state vector for Q-learning frameworks. This method provides a systematic approach to allow the FLHA and Q-learning to self-configure their parameters based on the historical data distribution. This can further improve the overall performance of the algorithm proposed in this paper.

To the best of our knowledge, this is the first study to implement hybrid Q-learning and subtractive clustering techniques to optimise the FLHA. This is also the first work to utilise this emerging fuzzy hybridisation system to directly trigger HO process rather than just tuning the HO parameters of the A3 event.

The remainder of this paper is organised as follows: Section 2 reviews the state-of-the-art literature in the fields of HO optimisation. Section 3 introduces the framework of conventional FLHA. Section 4 describes how to synergise subtractive clustering techniques and Q-learning frameworks to enhance the FLHA. The simulation environment and evaluation results are discussed in Section 5. Section 6 concludes the paper.

## 2. RELEVANT WORKS

### 2.1 Threshold based HO optimisation approaches

Several approaches related to the self-optimisation method can be found in the literature. References [10]–[12] achieved self-optimisation using threshold comparisons with specific metrics to optimise the parameters within the A3 event, that is, the time to trigger (TTT) and HO margin (HOM). These two parameters are adopted to avoid the ping-pong effect that causes noise and interference by adding an extra condition before HO triggering. Traditionally, these two parameters need to be frequently and manually adjusted by conducting massive measuring campaigns and statistical analysis. In [10], the authors proposed an auto-tuning algorithm that utilises user speed and RSRP as decision criteria to continuously tune the HOM and TTT in 5G-HetNets based on metaheuristic algorithms. In [11], the authors proposed an approach to categorise HO failures into three types: too late, too early, and wrong cell. The evaluated results will be compared to a predefined threshold to determine whether HOM and TTT need to be updated within a specific period. Similarly, the authors demonstrated an algorithm in [12] to detect ping-pong and fast-moving users by evaluating the users' dwelling and remaining service time within one cell. The evaluated results will be compared with predetermined thresholds to decide whether HOM and TTT need to adjust or transfer the UE's connection to the macrocell. The simulation results in [10]–[12] indicated that all proposed approaches could effectively enhance the user mobility robustness by reducing unnecessary HOs, ping-pong effects, and HO failures. However, these papers did not further discuss how to define each algorithm's threshold value. Therefore, the mobility robustness optimisation in [10]–[12] was not fully automated.

## 2.2 Reinforcement learning based HO optimisation approaches

To enable a fully automated and cognitive network, reinforcement learning has attracted considerable research attention. References [13]–[16] demonstrated the reinforcement learning-based self-optimisation function. In [13], the authors developed an HO detection mechanism to analyse the measured data from UE and calculated HO events to avoid false HOs. Moreover, the authors also proposed a SON mechanism from Markov's decision process (MDP). The two state variables are defined from radio states HOM and TTT. The HOM and TTT can be tuned simultaneously using the proposed mechanism to improve system performance. In [14], the average UE speed was considered in the Q-learning framework to select an appropriate time by tuning the HOM and TTT. Next, several decision criteria, that is, the RSRP, UE movement direction, and location were considered in a multiple attribute decision-making algorithm to select the HO target. In [15], the authors proposed a cognitive SON function based on a Q-learning framework. The proposed method can learn the optimal HOM and TTT for particular UE mobility patterns in the network. A cooperative learning strategy is applied during the Q-learning training stage to share experiences among cells and hence speed up the learning process. Based on a similar approach, another self-optimisation algorithm with a load-balancing objective was also developed in [15]. In [16], a distributed Q-learning algorithm was developed to learn the optimal BS selection method from each cellular network to achieve load balancing. Multiple attributes, including the channel load, HO duration, and signal to interference plus noise ratio (SINR), are considered as Q-learning reward functions. The simulation results in [13]–[16] showed that reinforcement learning is a powerful tool to enable automated network optimisation for either mobility robustness or mobility load optimisation. The HO performance, that is, the number of HOs, HO failure rate, and call drop rate, were effectivity reduced by these Q-learning-based self-optimisation mechanisms. However, Q-learning can only store a limited number of state-action pairs in the trained Q table to ensure its training efficiency and action selection accuracy. In other words, Q learning is only suitable for discrete environments with limited states and actions.

## 2.3 Fuzzy logic based HO optimisation approaches

On top of Q-learning, the fuzzy logic algorithm is also widely used to achieve self-optimisation function as shown in [17]–[21]. References [17] and [18] demonstrated a fuzzy logic-based self-optimisation method for the HOM and TTT. In contrast, [19]–[21] applied a non-conventional approach known as the FLHA, which can achieve both timely and flexible mobility robustness optimisation. The FLHA's main strategy considers various metrics as the fuzzy inference engine input to estimate the HO probability. Subsequently, the estimated HO probability is adopted as the critical factor (the HO factor) to trigger the HO process. The simulation results in [17]–[21] proved that fuzzy logic is a very useful tool to build self-optimisation functions. However, fuzzy logic-based approaches have not further elucidated how to design proper fuzzy inference engines. Fuzzy inference engines consist of a set of fuzzy rules and fuzzy sets that can process input parameters to output in linguistic terms. Traditionally, the design of fuzzy rules and fuzzy sets requires the manual tuning of human experts and experience to obtain desired outputs. Thus, the traditional FLHA is not the universal solution to mobility robustness optimisation.

Based on the previously described reviews and analyses, both Q-learning and fuzzy logic have their strengths and weaknesses. The following section will introduce an emerging fuzzy system that integrates the advantages of fuzzy inference systems and Q-learning as a joint effort in HO decision making, thereby enhancing the mobility robustness of UE.

## 3. FUZZY LOGIC HANDOVER ALGORITHM

As previously mentioned, the FLHA's main strategy is to estimate the HO probability by considering multiple input parameters in the fuzzy inference engine. The estimated HO probability, known as the HO factor, is used to trigger the HO process. As shown in Fig. 1, the FLHA's general architecture consists of three stages: fuzzification, inference engine, and defuzzification.

In the *fuzzification process* stage, the crispy input metrics are mapped into a degree of membership and translated into linguistics variables (low, medium, and high) through a predefined fuzzy membership function (MF). For input values $x \in \Re$ and $x \in \langle 0,1 \rangle$, the fuzzy membership function $\mu_{\tilde{S}}(x)$ describes the fuzzy set $\tilde{S}$ within a universal discourse $X$. The $\mu_{\tilde{S}}(x)$ value represents the degree of membership of $x$ in $\tilde{S}$. In this paper, the Gaussian membership function (GMF), shown in Fig. 2, is adopted due to its smoothness and concise notation.

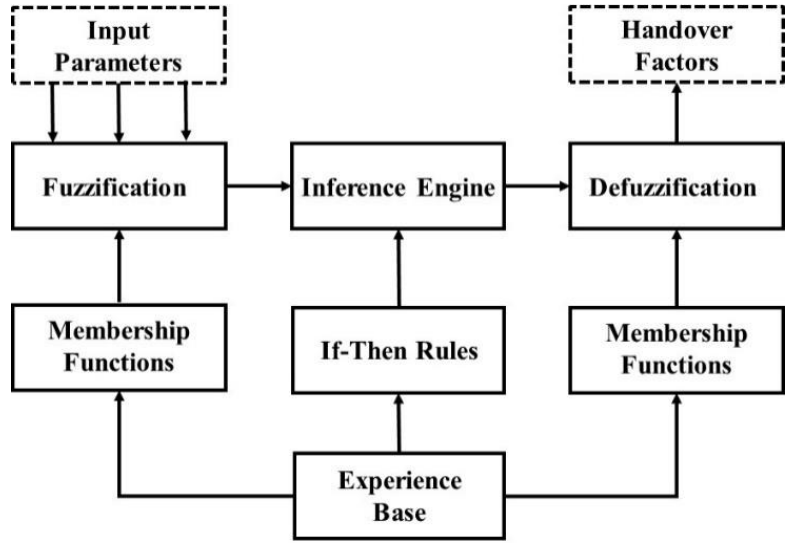

Figure 1. General architecture of the FLHA

The second stage *fuzzy inference engine* contains predefined fuzzy rules that link system inputs with output. An example of fuzzy rules in the FLHA with three input metrics ($x$, $y$, and $z$) and output HO factor ($w$) is expressed as

$$IF\ x_k == \tilde{S}_x^k\ and\ IF\ y_k == \tilde{S}_y^k\ and\ IF\ z_k == \tilde{S}_z^k\ THEN\ h_k = \widetilde{W}_h^k,\ for\ k = 1,2,\dots,n \quad (1)$$

where $\tilde{S}_x^k$, $\tilde{S}_y^k$, and $\tilde{S}_z^k$ are the fuzzy sets for the *k-th* input of metrics $x$, $y$, and z and $\widetilde{W}_h^k$ is the fuzzy set for the *k-th* output. Next, a max-min inference method is adopted to calculate the degree of membership for the output variables due to its computational simplicity [22]. A fuzzy implication operator is applied to obtain the consequent fuzzy process output from each fuzzy rule. The consequences of each fuzzy rule are then combined into a new fuzzy process output by a fuzzy aggregation operator. In other words, the rule with the highest degree of membership is chosen. For the *k-th* inputs to the FLHA, if there are $j$ fuzzy rules in the inference engine, the corresponding aggregated fuzzy process output is represented by a consequent MF $\mu_{\widetilde{W}_h^k}(h_k)$ as

$$\mu_{\widetilde{W}_h^k}(h) = \bigcup_k \left[\bigcap_{r=1}^{j}\left[\mu_{\tilde{S}_x^{k,r}}(x_k), \mu_{\tilde{S}_y^{k,r}}(y_k), \mu_{\tilde{S}_z^{k,r}}(z_k)\right]\right]\ for\ k = 1,2,\dots,n\ and\ r = 1,2,\dots,j \quad (2)$$

where $\mu_{\tilde{S}_x^{k,r}}(x_k), \mu_{\tilde{S}_y^{k,r}}(y_k)$, and $\mu_{\tilde{S}_z^{k,r}}(z_k)$ represent the consequent MF of the fuzzy process output by the *r-th* rule for the *k-th* input of metrics $x, y,$ and $z$.

The last stage of the *defuzzification process* is the opposite of fuzzification. In this stage, the aggregated fuzzy process output $\mu_{\widetilde{W}_h^k}(h)$ is converted into a crispy value from its centroid of the area as [22]

$$h_k = \frac{\int \mu_{\widetilde{W}_h^k}(h) \cdot h\, dh}{\int \mu_{\widetilde{W}_h^k}(h)\, dh} \quad (3)$$

where $h_k$ is the centroid of the area of aggregated fuzzy process output $\mu_{\widetilde{W}_h^k}(h)$, which is the crispy output value denoted as the HO factor in the FLHA. The HO factor is a string number from 0 to 1, where 0 represents the least liable to HO, and 1 refers to the most liable to trigger an HO process. The HO factor's value is implemented to determine the timing of HO triggering by comparing with one predefined threshold.

Based on the previously described analysis, the MF and fuzzy rules directly affect the output value and system performance. Typically, the MF design and fuzzy rules rely on expert experience and trial and error. More MFs ensure more accurate output. However, with the increasing number of input metrics and corresponding MFs, the design workload for fuzzy rules will also grow exponentially. For example, if there are $n$ FLHA inputs, each input has $j$ fuzzy sets, and thus, $j^n$ rules will need to be defined. When considering multiple parameters as the FLHA input, its design process becomes extremely complicated; consequently, it is difficult to ensure the system's reliability.

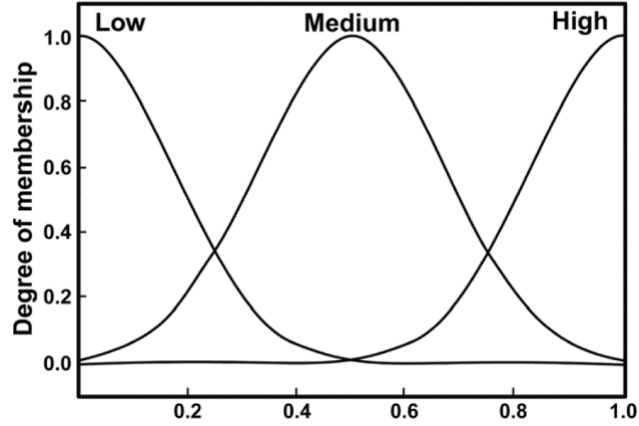
Figure 2. The Gaussian fuzzy membership function.

## 4. JOINT SELF-OPTIMISATION FLHA METHOD

To overcome the conventional FLHA's limitations, in this section, we integrate both subtractive clustering and Q-learning framework into the FLHA to give it self-configuration and self-optimisation functions. The proposed method's basic structure is described in Algorithm 1. Subtractive clustering is first applied to generate an MF for each metric based on the data distribution, which can ensure that all the FLHA's input metrics are correctly mapped into corresponding fuzzy sets (as detailed in Section 4.1). Second, the Q-learning framework is then implemented to learn the optimal policy as the fuzzy rules for the FLHA from the environment (as detailed in Section 4.2). The generated MF and fuzzy rules are then used to design the FLHA as the triggering mechanism. The output of the proposed method's HO factor is adopted to trigger the HO process (as detailed in Section 4.3).

| | *Algorithm 1: Main joint self-optimisation method for FLHA* |
|---|---|
| 1 | ***Self-configure GMF by subtractive clustering***:       (*Ref: Algorithm 2*) |
| 2 |   ***Input:*** *Historical data, that is, RSRP, SINR, and transmission distance* |
| 3 |   ***Output:*** *GMF for each input metric* |
| 4 | ***Self-optimise fuzzy-rules by Q-learning***       (*Ref: Algorithm 3*) |
| 5 |   ***Input:*** *State GMF from Algorithm 2, action, and reward function* |
| 6 |   ***Output****: Optimal fuzzy rules policy* |
| 7 | ***FLHA***       (*Ref: Section 2*) |
| 8 |   ***Input:*** *GMF from Algorithm 2, fuzzy rules from Algorithm 3* |
| |     *Normalised decision metrics: that is, RSRP, SINR, and transmission distance* |
| 9 |   ***Output:*** *HO factor* |
| 10 | ***HO triggering***       (*Ref: Algorithm 4*) |
| 11 |   ***Input:*** *HO factor* |
| 12 |   ***Output****: HO triggering decision* |
| 13 | ***End*** |

## 4.1 Self-configuring the MF using subtractive clustering

The GMF $\mu_{\tilde{s}}(x)$ is formulated as in Eq.4. Fig. 3 illustrates the physical meaning of each parameter.

$$\mu_{\tilde{s}}(x) = \left[1 + (\frac{x-v_i}{\sigma_i})^{b_i}\right]^{-1} \tag{4}$$

where $\tilde{s}$ is the fuzzy sets in the GMF, and each GMF includes $i$ fuzzy sets; $x$ is the normalised value of the input metrics. The value $\mu_{\tilde{s}}(x) \in (0,1)$ is the degree of membership of $x$ in $\tilde{s}$, $v_i$, $\sigma_i$, and $b_i$ represent the centre, width (standard deviation), and crossover slope of one fuzzy set, respectively. The slope at the crossover point (where $\mu_{\tilde{s}}(x) = 0.5$) is determined by $b_i$ and $\sigma_i$ as

$$slop_i = \frac{-b_i}{2\sigma_i} \tag{5}$$

The generalised GMF distributes fuzzy sets evenly at the axis, but this is suitable only if the data sets are uniformly sampled. Therefore, to ensure the effectiveness of the GMF and FLHA, the parameter in Eq.4 should follow the probability distribution of the input data sets. Due to lack of prior knowledge of the data distribution in most cases, the clustering technique is considered a reliable tool to extract characters from different data sets. In this paper, subtractive clustering is adopted due to its "one-pass" method that ultimately contributes to high computational efficiency. Motivated by [23], Algorithm 2 describes in detail how to generate the GMF for each input metric.

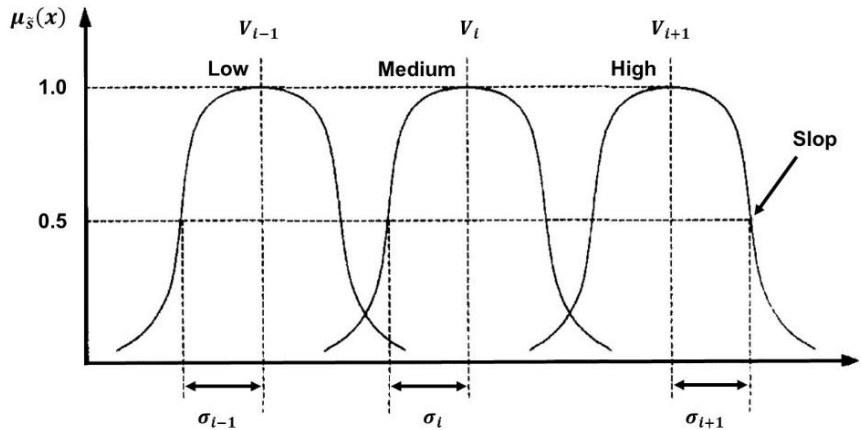

Figure 3. The physical meaning of the parameters in the generalised Gaussian fuzzy membership function (GMF).

As described in Algorithm 2, the input metrics are first normalised between 0 and 1 as input data set $x_{ij}$. Each metric represents one dimension of the input sets. In this paper, three metrics are used as the HO decision criteria, and m = 3. The core strategy of subtractive clustering is to find the data points with the highest potential using Eq.6. In Eq.6, $\alpha$ defines the influence range for neighbouring points. The data points outside this range have a limited influence on the potential calculation. After calculating each input data point's potential, the data point with the highest potential is selected as the first cluster. Next, the following cluster is located by updating the potential for other data points based on the previous cluster location as shown in Eq.8. $\beta$ is adopted to control the distance between each cluster. To avoid cluster centres that are too close to each other, we usually set $\frac{\alpha}{\beta} = 1.5$. When the condition in Eq.6 is met, the potential update is complete. $\varepsilon$ in Eq.6 is a small fraction known as the rejection ratio that determines the number of clusters and is inversely proportional to the number of clusters. The data points with the highest potential in each round of updating are selected as the following new cluster.

After obtaining all the clusters from each input set, the centre of cluster $\{\vec{x_i}^* = (x_{i1}, x_{i2}, ..., x_{im})\}$ for each input metric is adopted as the centre $\{v = (v_1, v_2, ..., v_i)\}$ of each fuzzy set. Eq.9 is used to define the width of the fuzzy sets, which is mainly dominated by the maximum and minimum values of the data set in the $j$-th dimension ($U_{j\text{-}min}$ and $U_{j\text{-}max}$). $\delta$ in Eq.9 is typically set between 2 and 3. The GMF's crossover slope is defined when the adjacent membership functions overlap by approximately 25%. The subtractive clustering parameters are set as $\{\alpha = 16, \beta = 12, \varepsilon = 0.005, \text{and } \delta = \sqrt{8}\}$ to define a suitable number of clusters for each fuzzy set in this paper. The designed GMF is also compared with the probability density function (PDF) of the input data sets for validation.

---

***Algorithm 2**: Self-configuring the GMF for each input metric*

1. **Input:** *Historical decision metrics data, that is, the RSRP, SINR, and transmission distance*
2. *Import normalised input data set:* $x_{ij}$, $i = 1, 2, ..., n$; $j = 1, 2, ..., m$
   $n$ = *number of data points*; $m$ = *number of dimensions of data sets*
3. *Define values for* $\alpha$, $\beta$, $\varepsilon$, *and* $\delta$
4. *Calculate* $U_{j\text{-}min}$ *and* $U_{j\text{-}max}$ *from the input data sets*
5. *Calculate the potential value of each data point*

$$P_i = \sum_{k=1}^{n} e^{-\alpha \|x_i - x_k\|^2} \quad i = 1, 2, \cdots, n; i \neq k \tag{6}$$

6. *Select the first centre* $x_1^*$ *with the highest potential* $P_1^*$
7. **While** $(P_k^* > \varepsilon P_1^*)$ \hfill (7)
8. \quad *Update the data point potential*
9. \quad $P_i \leftarrow P_i - P_k^* e^{-\beta \|x_i - x_k^*\|^2}$ \hfill (8)
10. \quad *Select the k-th centre* $x_k^*$ *with the highest potential* $P_k^*$
11. **End while**
12. *Determine the centres of the fuzzy sets within the GMF:* $x_1^*, ..., x_k^*$
13. *Determine the width of the fuzzy sets within the GMF:*

$$\sigma_i = \beta * \frac{(U_{j,max} - U_{j,min})}{\delta} \tag{9}$$

14. **End**
15. **Output:** *GMF of each input metric*

---

### 4.2 Self-optimising fuzzy rules using the Q-learning framework

Q-learning is a model-free and off-policy reinforcement learning approach that uses the temporal difference (TD) based on a set of Markov decision processes. The basic framework of the Q-learning-based FLHA is illustrated in Fig. 4. It consists of state ($\mathcal{S}$), action ($\mathcal{A}$), reward ($\mathcal{R}$), agent, and environment. At each time step $t$, the agent performs an action $a_t \in \mathcal{A}$ and undergoes a transition from state $s_t \in \mathcal{S}$ to $s_{t+1} \in \mathcal{S}$. Subsequently, the reward $r_t \in \mathcal{R}$ is provided to the agent by the environment. The agent's main objective is to learn the policy function $\pi$ that can select the optimal action at each state to maximise the accumulated reward in the long term.

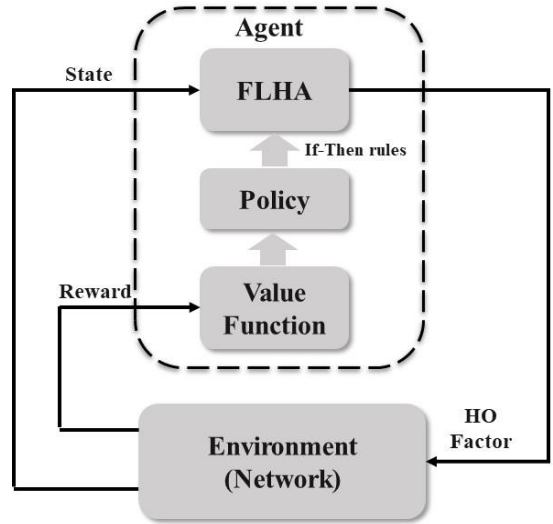

Figure 4. Basic Q-learning framework.

In this paper, to obtain the FLHA's optimal fuzzy rules, we define the FLHA as the Q-learning agent. At time step $t$ and area $l$, the input decision metrics are first normalised between 0 and 1 and then mapped into the GMF (defined in Section 3.1) to find its fuzzy set. The states $s_{i,l,t} \in \mathcal{S}$ for UE $i$ is represented by the combination of the fuzzy sets as

$$s_{i,l,t} = \{\widetilde{s_{RSRP}^{i,l,t}}, \widetilde{s_{SINR}^{i,l,t}}, \widetilde{s_d^{i,l,t}}\}, \tag{10}$$

where $\widetilde{s_{RSRP}^{i,l,t}}$, $\widetilde{s_{SINR}^{i,l,t}}$, and $\widetilde{s_d^{i,l,t}}$ represent the fuzzy sets of the RSRP, SINR, and transmission distance ($d$), respectively. If a normalised value is located adjacent to different memberships, the membership with a higher grade of $x$ is the used to represent its fuzzy set. For example, assume Fig. 2 is the MF for RSRP, SINR, and d. If the normalised input metrics for these three inputs equal to 0.2, 0.4, and 0.8, the corresponding fuzzy sets are mapped as *low*, *medium*, and *high*, respectively. Therefore, the state vector $s_{i,l,t}$ is represented as

$$s_{i,l,t} = \{RSRP(low), SINR(medium), d(high)\} \tag{11}$$

If the HO process is triggered at time $t$, the state at $t+1$ is updated based on the RSRP, SINR, and $d$ from new serving BS of the UE. Otherwise, the state at $t+1$ is updated based on the input metric of the current BS.

At time step $t$ and area $l$, the actions $a_{i,l,t} \in \mathcal{A}$ for UE $i$ are defined as

$$a_{i,l,t} = \begin{cases} a_1 = trigger\ HO\ process \\ a_2 = maintain\ curren\ connection \end{cases} a_1, a_2 \in \mathcal{A}, \tag{12}$$

If the agent chooses action $a_1$ to execute, the handover process will be triggered by UE. The UE's connection will be switched to a neighbouring BS with the highest SINR at time step $t+1$. Otherwise, the UE will maintain its connection with the current serving BS.

At time step $t$ and area $l$, after the agent performs an action $a_{i,l,t} \in \mathcal{A}$ for UE $i$, the agent's reward $r_{i,l,t} \in \mathcal{R}$ is defined as

$$r_{i,l,t} = v\left(\widetilde{s_{i,l,t+1}^{RSRP}}\right) + v\left(\widetilde{s_{i,l,t+1}^{SINR}}\right) + v\left(\widetilde{s_{i,l,t+1}^d}\right), \tag{13}$$

where $v\left(\widetilde{s_{i,l,t+1}^{RSRP}}\right)$, $v\left(\widetilde{s_{i,l,t+1}^{SINR}}\right)$, and $v\left(\widetilde{s_{i,l,t+1}^d}\right)$ represent the centre value of the fuzzy set $\widetilde{s_{i,l,t+1}^{RSRP}}$, $\widetilde{s_{i,l,t+1}^{SINR}}$, and $\widetilde{s_{i,l,t+1}^d}$, respectively. If the FLHA executes the HO process at time $t$, the reward signal is obtained from the new serving BS. Otherwise, the reward signal is received from the current serving BS.

After establishing $<\mathcal{S}, \mathcal{A}, \mathcal{R}>$ in the Q-learning framework, a value function $Q(s,a)$, also known as the Q-value, is defined to represent the value of a state-action pair. In other words, $Q(s,a)$ indicates the expected cumulative reward obtained when performing action $a$ at state $s$. A policy function $\pi$ is adopted to decide which action needs to be performed in each state. The value function following policy $\pi$ is formulated as

$$Q^\pi(s,a) = E_\pi\{\mathcal{R}_t | s_t = s, a_t = a\}$$
$$= E_\pi\left\{\sum_{k=0}^{\infty} \gamma^k r_{t+k+1} \middle| s_t = s, a_t = a\right\}, \tag{14}$$

where $E_\pi\{\}$ is the expected value under policy $\pi$ and $\gamma \in (0,1)$ is adopted as a discount factor to determine the relative importance of future rewards. During the Q-learning training stage, the agent approximates the optimal value function $Q^*(s,a)$ received by a TD error that describes the difference between the actual and estimated Q-values. The updated Q-value is formulated as

$$Q(s_t, a_t) \leftarrow Q(s_t, a_t) + \alpha\left[r_{t+1} + \gamma \max_a Q(s_{t+1}, a) - Q(s_t, a_t)\right] \tag{15}$$

$\alpha \in (0,1)$ is the learning rate to balance the latest and previous knowledge. $\epsilon$-greedy is then adopted to control the trade-off between exploration and exploitation of the state-action space. With $\epsilon$-greedy, at time step t, the agent performs optimal action $a(s) = \arg\max_k Q(s,k)$ with probability $1 - \epsilon$; otherwise, a random action is performed. When $\epsilon=0$, the action with the highest Q-value is always performed.

A table (known as Q-table) with the Q-value of each state-action pair can be obtained after the learning process. Since two actions are defined for each state, the trained Q-table must have three columns (one for state, two for actions). Then the difference of the Q value corresponding to each state can be calculated as,

$$\Delta Q = Q(s, a_1) - Q(s, a_2) \tag{16}$$

Where, a state with the positive $\Delta Q$ means the action "*trigger HO process*" can obtain more benefit than "*maintain current connection*". As such, a higher $\Delta Q$ means HO is in the higher probability to be triggered. Conversely, a lower $\Delta Q$ indicates HO is less likely to be triggered.

The $\Delta Q$ in each state will be normalised and applied to subtractive clustering (Algorithm2) to locate its cluster, and subsequently generate output GMF for the HO factor. Each cluster of $\Delta Q$ can represent one fuzzy set of output GMF. Each state in Q-table can generate one fuzzy rule based on the cluster of $\Delta Q$. For example, if there are 4 clusters can be located for $\Delta Q$, which can be denoted as HO process is in *very high / high / low / very low probabilities to trigger* respectively based on their centre value in descending order. For a state as Eq (11), if their $\Delta Q$ belong to *very high*, a fuzzy rule can hence be generated as,

$$IF\ RSRP == low\ and\ IF\ SINR == medium\ and\ IF\ d == high\ THEN\ HO\ factor = high$$

Based on this method, each state can generate one fuzzy rule for the proposed FLHA system. The algorithm of self-optimised FLHA based on Q-learning is described as algorithm 3.

| | *Algorithm 3: Self-optimisation FLHA for UE i in area l* |
|---|---|
| 1 | **Input:** *historical data, i.e. RSRP, SINR, d etc.* |
| 2 | *Generated GMF for each input metric as Algorithm 2* |
| 3 | *Initialise Q(s,a) arbitrarily, $\forall\ s\in\mathcal{S}, a\in\mathcal{A}$ and Q(terminal_state.)=0* |
| 4 | **for** *each epoch* **do** |
| 5 |    *Initialise $\mathcal{S}$ from GMF* |
| 6 |    **for** *each time step t* **do** |
| 7 |      *Choose $a_{i,t}\in\mathcal{A}$ from $\mathcal{S}$ using $\epsilon$ -greedy policy* |
| 8 |        *if $a_{i,t}$ = trigger HO process* |
| 9 |          *Execute the HO process* |
| 10 |          *Select neighbouring BS with max(SINR) as the HO target $BS_{j+1}$* |
| 11 |          *Transfer UE's connection to new $BS_{j+1}$, observe reward from $BS_{j+1}$* |
| 12 |        *else* |
| 13 |          *Maintain UE's connection with current $BS_j$, observe reward from $BS_j$* |
| 14 |        *end if* |
| 15 |      *Update Q-value by Eq(15)* |
| 16 |      $\mathcal{S} \leftarrow \mathcal{S}$ |
| 17 |      *Until $\mathcal{S}$ is terminal* |
| 18 |    *end* |
| 19 | *end* |
| 20 | *Output: Q-table* |
| 21 | **for** *each state in Q-table* **do** |
| 22 |    *Calculate $\Delta Q$ by Eq(16)* |
| 23 | *end* |
| 24 | *Apply $\Delta Q$ to Algorithm 2* |
| 25 | **Output:** *Cluster for $\Delta Q$, output GMF for HO factor* |
| 26 | *Generate fuzzy rule at each state* |
| 27 | **Output:** *Fuzzy rules, GMF of HO factor* |

**4.3 HO triggering by the self-optimised FLHA**

Subtractive clustering and the Q-learning framework can be integrated to establish the joint self-optimised FLHA, which is deployed at the UE as a triggering mechanism. During the UE movement, the collected HO-related metrics RSRP, SINR, and UE transmission distance are first normalised between 0 and 1 and utilised as input for the joint self-optimised FLHA to obtain the HO factor. The HO process is then triggered if the HO factor is higher than the threshold and, subsequently, the HO event is sent by the UE to its serving BS. The serving BS will then execute the following HO procedure and switch the UE's connection to a neighbouring BS with the highest SINR. If the HO factor is smaller than the threshold, the UE will then maintain its connection with its current serving BS.

The HO triggering threshold is also defined from the trained Q-table, a state with $\Delta Q$ approximately equal to zero will first be selected. This state is a critical point where the same benefit can be received by triggering a handover or maintaining the connection. Afterwards, the centre value of the fuzzy sets in this state will be implemented into the proposed FLHA to obtain a crispy value as HO triggering threshold.

The triggering process of self-optimised FLHA is shown in algorithm 4.

| | *Algorithm 4: HO triggering by joint self-optimisation FLHA* |
|---|---|
| 1 | *While(true)* |
| 2 | *Send Measurement_Report* |
| 3 | ***Input:*** *RSRP, SINR, d etc. from serving and neighbouring BS* |
| 4 | *GMF from subtractive clustering, If-Then rules form Q-learning* |
| 6 | ***Output:*** *Defuzzification = HO factor* |
| 7 | *if HO factor > threshold* |
| 8 | *Select neighbouring BS with max(SINR) as the HO target $BS_{j+1}$* |
| 9 | *Send HANDOVER_REQUEST* |
| 10 | *Send Path_Switch_Request* |
| 11 | *Transfer UE's connection to the HO target $BS_{j+1}$* |
| 12 | *else* |
| 13 | *Maintain UE's connection with current $BS_j$* |
| 14 | *end if* |
| 15 | *end* |

## 5. PERFORMANCE ANALYSIS

**5.1 Analysis design**

In this study, a 1000 m × 1000 m two-tier HetNets scenario that compromises two LTE macro BSs and 16 5G small BSs is developed using MATLAB as shown in Fig.5 to evaluate the performance of the proposed triggering mechanism. All the BSs are uniformly distributed over the geographical area with a distance of approximately 350 m. The 4G macro BSs operate at a frequency band at 1.5-2 GHz, and the 5G small BSs work at the mm-wave band. The Urban Macro (UMa) and Urban Micro (UMi) propagation models in Reference [24] are adopted to model the channel of the macro and small cells. The additive white Gaussian noise (AWGN) and Rayleigh noise are added to the channel as noise. There are 40 UEs randomly moving in the proposed environment with constant speeds of 30, 75, and 120 km/h to evaluate the mobility robustness of the proposed method in the range of low, medium, and high speeds. The detailed simulation parameters are shown in Table 1.

Each time step in the simulation includes updating the UE's position, propagation calculation, and HO decision-making using the different triggering mechanisms. The conventional RSRP-based triggering mechanism in the A3 event [6] and the experience-based conventional FLHA are adopted as the baselines and competitive algorithms respectively to compare with the proposed approach. To evaluate the effectiveness of subtractive clustering, the FLHA that is only optimised by Q-learning and generalised GMF (FLHA-Q) is also adopted as a competitive algorithm.

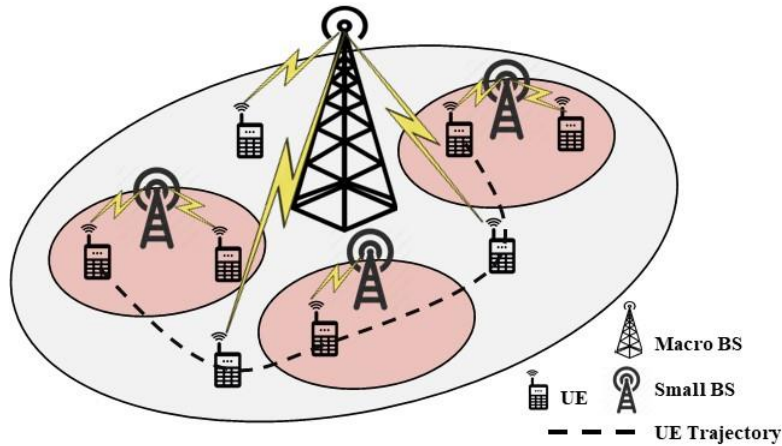

Figure 5. HetNets simulation environment.

Table 1. Simulation parameters

| Parameters | Specification | |
|---|---|---|
| | Macro BS | Small BS |
| Carrier frequency (GHz) | 1.5 ~ 2 | 28 |
| Subcarrier spacing (KHz) | 15 | 30 |
| System bandwidth (MHz) | 20 | 100 |
| Physical resource block | 100 | 275 |
| Number of BSs | 2 | 16 |
| Subcarriers per PRB | 12 | |
| BS transmitted power (dBm) | 40 ~ 45 | |
| Duration of simulation | 10000 s | |
| Mobility model | Random direction | |
| Number of UE | 40 | |
| UE speed (km/h) | 30, 75, 120 | |
| Type of Noise | AWGN, Rayleigh | |
| HO preparation time (ms) | 10ms | |
| HO execution time (ms) | 10ms | |

### 5.2 Key performance indicators (KPIs)

To evaluate the UE's mobility robustness under the different algorithms, three mobility-related KPIs are adopted in this paper: the HO, ping-pong HO, and HO failure ratios. Besides, the proposed algorithm should minimise these three ratios while maintaining other KPIs at a high level to achieve Pareto optimisation. The network latency and user throughputs are also implemented as KPIs to show the proposed algorithm's effectiveness.

The HO ratio ($\overline{HOR}$) is also known as the HO probabilities that measure how the HO process is frequently triggered by the HO triggering mechanism. The average $\overline{HOR}$ in each time step per user is measured as

$$\overline{HOR} = \frac{\sum_{j=1}^{N_u} NOH}{N_u \times T}, \tag{17}$$

where $NOH$ represents the total number of HOs in each UE throughout the simulation, $N_u$ is the number of UEs in the environment, and $T$ is the simulation duration.

The second KPI, the ping-pong HO ratio ($\overline{PPHO}$), measures the occurrence of unnecessary HOs between two BSs. A ping-pong HO is counted when UEs continually HO between two BSs in a certain interval $T_p$. Therefore, the average ping-pong HO ratio per UE is calculated as

$$\overline{PPHO} = \frac{N_{PPHO}}{NOH}, \tag{18}$$

where $N_{PPHO}$ is the total number of ping-pong HOs counted per UE throughout the simulation.

The third KPI, the HO failure ratio, measures the reliability of the proposed HO triggering mechanism. If the HO process is triggered too early, too late, or switches to the wrong cell, the entire HO procedure may fail to complete. The average HO failure ratio ($\overline{HOF}$) per UE is calculated as

$$\overline{HOF} = \frac{N_{HOF}}{NOH}, \tag{19}$$

where $N_{HOF}$ is the number of HO failures occurring per UE throughout the simulation.

The sum throughput at network KPI evaluates the quality of the user experience. The sum throughput ($\Gamma_{total}$) throughout the simulation is measured by Shannon's capacity theory, which is described as

$$\Gamma_{total} = B \times (log_2(1 + 10^{\gamma_{j,i}/10})), \tag{20}$$

where B is the bandwidth assigned to users and $\gamma_{j,i}$ is the SINR between UE $i$ and BS $j$.

The last KPI, network latency, reflects the quality of the user experience. Based on the analysis in [25], the network latency ($\hat{\Delta}_{i,j}^t$) between UE $i$ and BS $j$ at time $t$ is expressed as

$$\hat{\Delta}_{i,j}^t = \frac{\theta}{r_i} + \ell_{edge} \times \frac{d_{i,j}}{d_y} + \ell_{ho}, \tag{21}$$

where $\theta$ is the size of a packet transmitting at the channel and $r_i$ is the data rate of UE $i$. Therefore, the first part of Eq.(21) calculates the transmission latency; $\ell_{edge}$ is the maximum propagation latency when the UE is at the edge of cell coverage $d_y$. $d_{i,j}$ is the transmission distance between the UE and BS. The second part of Eq.(21) obtains the propagation latency.

The last part of Eq.(21) measures the HO latency, which is the interval from the execution of the HO process to its completion.

The performance gain between the proposed and competitive approach is measured as,

$$\Delta KPI = \frac{KPI_p - KPI_c}{KPI_p} \tag{22}$$

where $KPI_p$ and $KPI_c$ represents the KPI obtained by the proposed and competitive approach respectively.

### 5.3 Simulation results

Fig. 6 depicts the GMF generated by subtractive clustering for each input metric. Fig.7 (a)-(c) shows each input metric's PDF with 40000 input data points. Fig.7 (d) is the PDF of the HO factor with 30 input data. An apparent relationship is found between the GMF and the corresponding PDF. The distribution of RSRP and SINR in Fig.6 nearly follows the Gaussian distribution with a mean value of approximately 0.4. Thus, the GMF of the RSRP and SINR are correspondingly concentrated at 0.4. The PDF of $d$ has no apparent concentration and is almost evenly distributed between 0 and 1. This distribution is due to the UE moving entirely randomly within the simulation scenario, and thus the data distribution of $d$ is uniform. However, $d$ is a cost criterion in HO decision-making, as a lower $d$ can contribute to lower latency and better radio state. Consequently, the GMF of $d$ is almost evenly separated between 0.3 and 1 with a reverse linguistic expression to the other two metrics. The range between 0 and 0.3 in normalised $d$ means the UE has a very long transmission distance, which is impossible as the HO must be triggered in this range. Accordingly, the GMF of $d$ without a fuzzy set is between 0 and 0.3.

The results in Fig. 6 and Fig. 7 show that subtractive clustering effectively extracts the features of the input set and generate the GMF based on their features accordingly. By implementing this method to create the GMF for the FLHA, the subjective error during the MF design can thus be eliminated. Moreover, this method can more accurately map the input data into the corresponding input sets, which can further enhance the FLHA's performance. In practice, adopting subtractive clustering can minimise the proposed algorithm's maintenance and optimisation capital, as it allows the algorithm to self-configure its parameters from historical data.

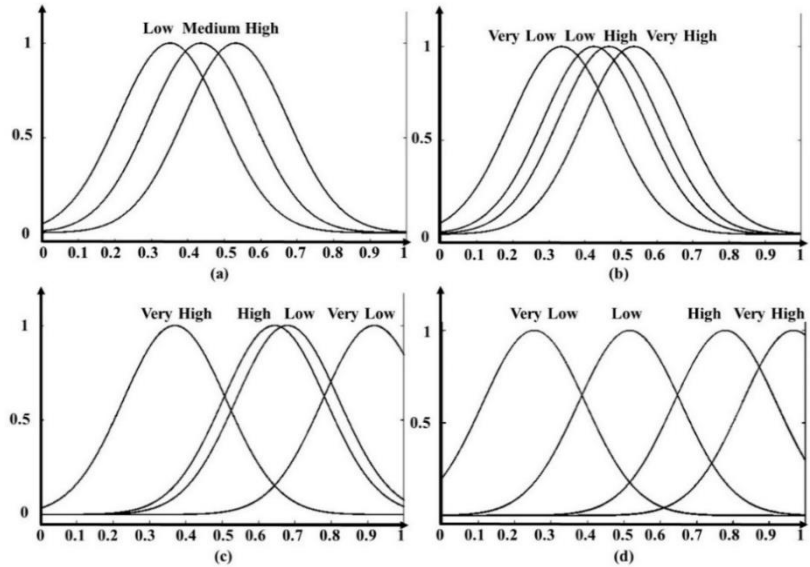

Figure 6. GMF generated by subtractive clustering for each input metric (a)RSRP, (b)SINR, (c)d and (d)HO factor

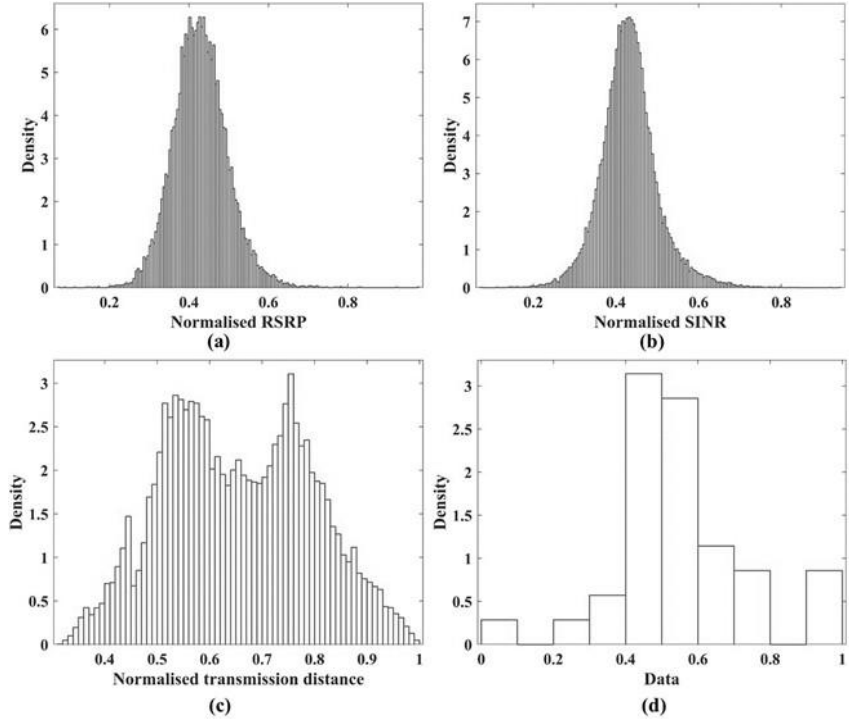

Figure 7. PDF of each input data (a)RSRP, (b)SINR, (c)d and (d)HO factor

Table 2. Fuzzy rules generated by the Q-learning framework.

| Rule No. | RSRP | SINR | d | ΔQ | HO factor |
|---|---|---|---|---|---|
| 1 | high | very high | very low | -11.020171 | very low |
| 2 | low | very high | very high | -10.967592 | very low |
| 3 | medium | very high | very low | -9.45235 | very low |
| 4 | low | very high | very low | -7.474153 | very low |
| 5 | high | low | high | -2.291506 | low |
| 6 | high | low | very high | -1.858905 | low |
| 7 | medium | very high | low | -1.855775 | low |
| 8 | high | very high | low | -1.58068 | low |
| 9 | medium | high | low | -0.584645 | low |
| 10 | medium | high | high | -0.555311 | low |
| 11 | high | high | low | -0.518546 | low |
| 12 | high | high | high | -0.407398 | low |
| 13 | low | very high | low | -0.386932 | low |
| 14 | medium | very high | high | -0.378761 | low |
| 15 | low | high | low | -0.335239 | low |
| 16 | low | high | high | -0.306926 | low |
| 17 | low | very high | high | -0.209441 | low |
| 18 | medium | low | high | -0.186788 | low |
| 19 | low | low | high | -0.119605 | low |
| 20 | high | very high | high | -0.093618 | low |
| 21 | medium | low | very high | 0.100416 | high |
| 22 | medium | high | very high | 0.122922 | high |
| 23 | low | high | very high | 0.210386 | high |
| 24 | medium | very low | very high | 0.25269 | high |
| 25 | low | low | very high | 0.447754 | high |
| 26 | high | high | very high | 0.608175 | high |
| 27 | low | very low | very high | 2.620136 | very high |
| 28 | high | low | low | 5.310236 | very high |
| 29 | medium | very high | very high | 7.362421 | very high |
| 30 | high | very low | very high | 8.592641 | very high |

In conventional FLHA, the design of the rules relies on prior experience. For an FLHA with three input metrics, if each input has four fuzzy sets, there are $C_4^1 \times C_4^1 \times C_4^1 = 64$ rules that need to be defined. Moreover, the FLHA deployed in varying application scenarios may use different rules. Thus, it is impossible to define optimal fuzzy rules for the FLHA in different scenarios. In this study, we adopt the Q-learning framework to learn the optimal policy from the environment as the FLHA's fuzzy rule. Table 2 illustrates the fuzzy rules generated by the Q-learning framework.

As shown in Table 2, 30 rules are generated based on Q-learning and the GMF from subtractive clustering. Theoretically, based on the GMF in Fig. 6, $C_3^1 \times C_4^1 \times C_4^1 = 48$ rules can be defined for three inputs. However, in real situations, some combinations of the fuzzy rule cannot exist as each metric may conflict with one another. The adoption of Q-learning in the FLHA allows the FLHA to self-configure and self-optimise its fuzzy rules by interacting with the environment rather than experience. This feature could eliminate the effect of subjective error in the FLHA and minimise the maintenance and optimisation capital of the proposed algorithm.

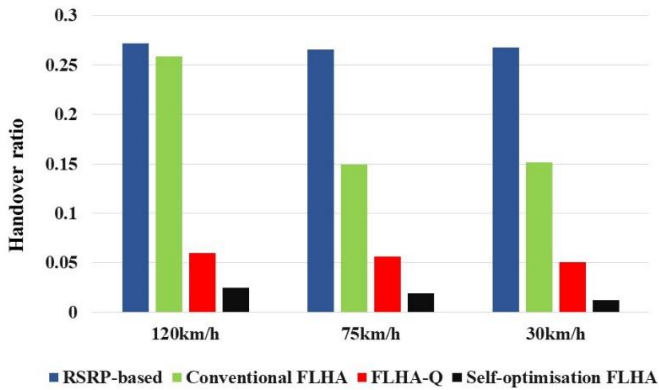

Figure 8. Average HO ratio versus different speeds.

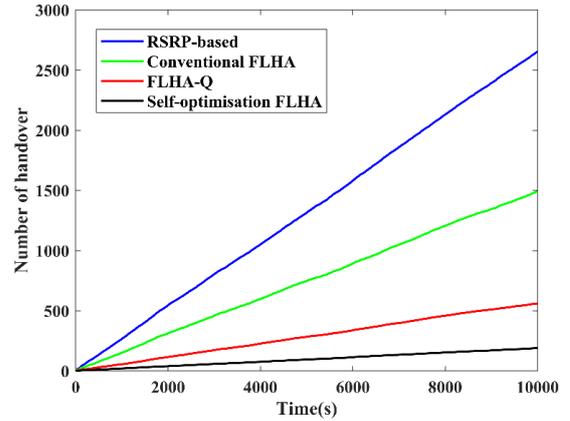

Figure 9. Number of HOs versus simulation times.

Figs. 8 and 9 illustrate how frequently the HO process is triggered using different approaches. The simulation results in Fig.8 show that as the UE speed increases, the HO ratio decreases under all approaches. This occurs because the moving range of low-speed UEs is smaller than that of high-speed UEs, and thus, fewer HOs are triggered. The simulation results also show that the proposed self-optimised algorithm can significantly reduce the HO ratio and the number of HOs for all the speed scenarios than the other algorithms. The overall HO ratio under self-optimisation FLHA is only around 0.02, and its performance is approximately 5-10 times higher than that of the baseline (RSRP) and conventional FLHA based approaches. Moreover, the adoption of the GMF generated by subtractive clustering in the FLHA can further reduce HO ratios from 0.05 to 0.02 compared with the Q-learning FLHA using the generalised GMF.

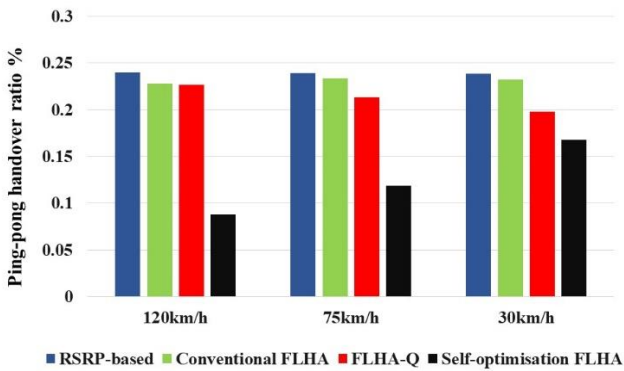

Figure 10. Ping-pong HO ratio versus different speeds.

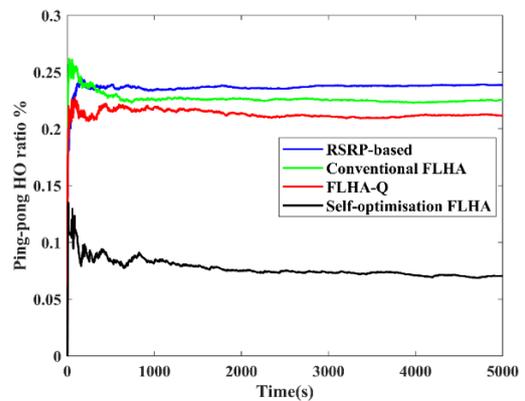

Figure 11. Ping-pong HO ratio versus simulation times.

Figs. 10 and 11 show the ping-pong HO ratios under the different approaches. The results in Fig.10 demonstrate that the speed has a limited effect on the ping-pong HO for the conventional RSRP, conventional FLHA, and Q-FLHA. The ping-pong ratio of the self-optimised FLHA increases with the decrease of UE speed. However, the ping-pong HO ratio of self-optimised FLHA still outperforms the other approaches among all the speed scenarios. The proposed approach can reduce the ping-pong HO ratio to 0.07%, 0.12% and 0.16% at three speed respectively. As measured by Eq.22, the proposed method's performance concerning ping-pong handover ratios is approximately 2.4 times higher than the other three competitive approaches.

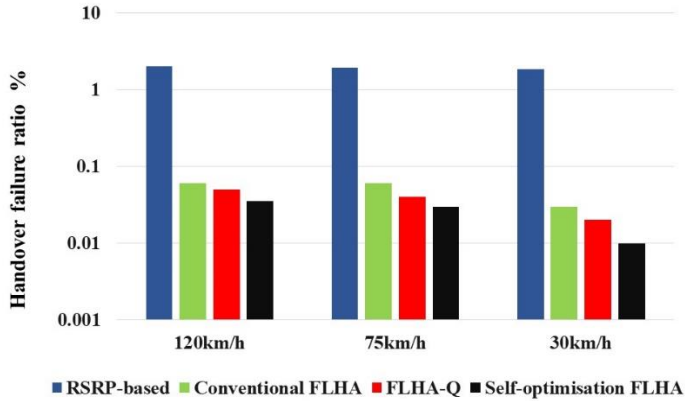 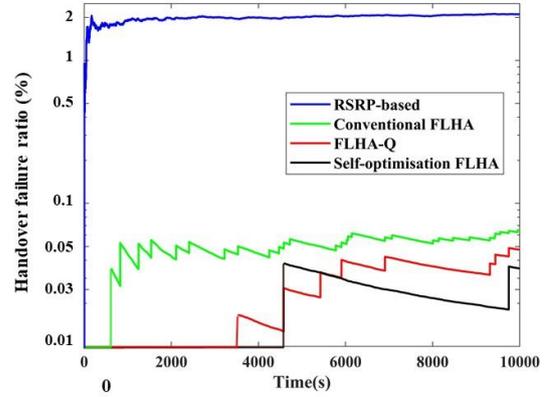

Figure 12. HO failure rate versus different speeds (y axis in log scale).   Figure 13. HO failure rate versus simulation times (y axis in log scale).

Figs. 12 and 13 show the HO failure ratio using the different approaches. The proposed method can reduce the HO failure ratio to 0.035%, 0.03% and 0.01% at three speed respectively. As evaluated by Eq.22, the self-optimised FLHA can achieve 1-2 times gain in handover failure when compared with the conventional FLHA and Q-FLHA.

The results demonstrate that all the FLHA-based triggering methods can achieve near-zero HO failure rates under all the speed scenarios. This result is due to HO failure that is mainly related to the SINR of the UE. One of the fuzzy rules defined in the FLHA is that if the RSRP is at a very low level, then the HO probability is high. Based on this rule, when the SINR of the UE is considered at a very low level in the FLHA, the FLHA will subsequently execute the HO process. Of note, this rule was first discovered by professionals and then applied to network maintenance and the FLHA. However, by adopting the Q-learning framework, the agent also can learn this rule by interacting with the environment. This demonstrates the powerful learning ability of Q-learning, which can effectively generate the optimal fuzzy rule for the FLHA.

According to the analysis above, the improvement of these three KPIs achieved by the self-optimisation FLHA are only around 0.1 % ~ 2%, which seems to be minimal. However, since 5G needs to provide millions of connections per kilometre square, this improvement could greatly ensure seamless mobility signalling, thus improving the user experiences and reducing a lot of unnecessary energy consumption.

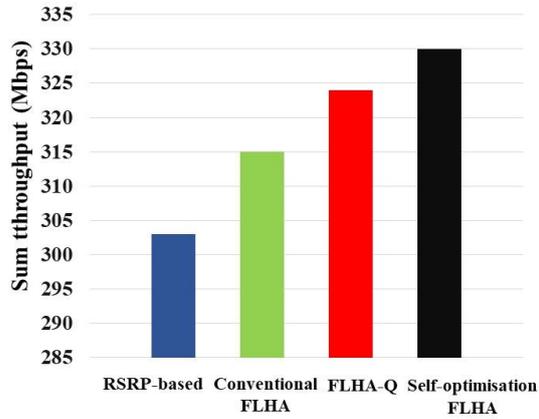 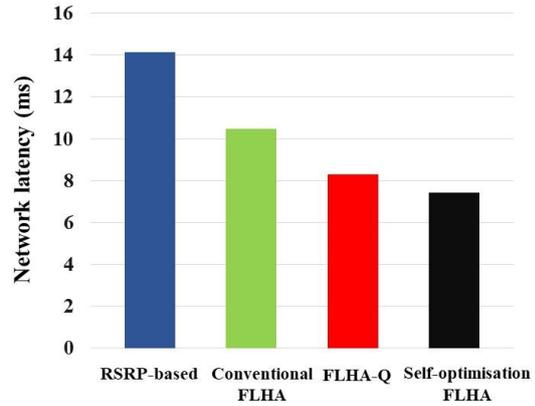

Figure 14. Sum throughput using the different approaches.    Figure 15. The average network latency using the different approaches.

The KPIs in Figs. 14 and 15 are utilised to evaluate the quality of the user experience. Some existing approaches focus only on improving mobility robustness but degrade the other KPIs related to load balancing and the user experience. To achieve Pareto optimisation in the network, the proposed algorithm should not only improve the mobility robustness of the user, but also maintain the other indicators at a high level.

Fig. 14 shows the sum network throughput under different approaches. The simulation results indicate that the proposed self-optimised FLHA can increase throughput by approximately 8%, 4.7%, and 1.9% compared with the conventional RSRP, conventional FLHA, and FLHA-Q, respectively.

Fig. 15 shows the average network latency using different approaches. The proposed self-optimised FLHA has the lowest latency. This occurs because the self-optimised FLHA can lead to a lower ping-pong HO ratio and higher throughput, which can effectively reduce HO and transmission latency. However, as the transmission distance is also one of the decision criteria of the self-optimised FLHA, it can result in a low propagation latency. The simulation results indicate that the proposed algorithm can decrease the overall latency by 47%, 27%, and 3.7% compared with the conventional RSRP, conventional FLHA, and FLHA-Q, respectively.

Based on the previously described analysis, the proposed self-optimised FLHA outperforms the other three algorithms in all the evaluated KPIs and speed scenarios. This strength may be due to the following reasons: first, the FLHA can make decisions in uncertain environments by compromising multiple conflicting input metrics. The GMF of the FLHA can also minimise the impact of interference and noise in decision-making. Second, the GMF generated by subtractive clustering can adequately reflect the distribution of the input data. This feature can more accurately map the input metrics of the FLHA into the corresponding fuzzy set, increasing the decision-making accuracy. Finally, the adoption of Q-learning can learn the optimal policy from the environment as fuzzy rules. The optimal fuzzy rules allow the FLHA to make more precise HO decisions based on environmental changes.

## 6. CONCLUSION

In order to enhance the mobility robustness of the user and reduce the network maintenance capital in 5G-HetNets, this paper proposed a self-optimised FLHA from the SON concept. The proposed approach integrated both Q-learning frameworks and subtractive clustering into the conventional FLHA to empower algorithms with SON functionality. Subtractive clustering can generate the GMF from historical data to enable the FLHA self-configure its MF. Q-learning can also learn the optimal HO policy from the environment to allow the FLHA to self-optimise its fuzzy rules. The simulation results indicated that the proposed self-optimised FLHA could enhance UE mobility robustness by significantly reducing the HO, ping-pong HO, and HO failure ratios while maintaining the network throughput and latency KPIs at a high level. Moreover, SON functionality can also minimise the proposed algorithm's maintenance and optimisation capital in realistic environments.

## ACKNOWLEDGEMENTS

The authors acknowledge financial support from the International Doctoral Innovation Centre (IDIC), Ningbo Education Bureau, Ningbo Science and Technology Bureau, and the University of Nottingham.